\documentclass[aps,prc,twocolumn,groupedaddress,showpacs]{revtex4}
\usepackage{graphicx}

\begin{document}

\title{Symmetry energy and the isoscaling properties of the fragments produced in $^{40}$Ar, $^{40}$Ca + $^{58}$Fe,
            $^{58}$Ni reactions at 25 $-$ 53 MeV/nucleon}
\author{J. Iglio}
\author{D.V. Shetty}
\altaffiliation{Corresponding author}
\author{S.J. Yennello}
\author{G.A. Souliotis}
\author{M. Jandel}
\author{A. Keksis}
\author{S. Soisson}
\author{B. Stein}
\author{S. Wuenschel}
\affiliation{Cyclotron Institute, Texas A\&M University, College Station, TX 77843}
\author{A.S. Botvina}
\affiliation{Institute for Nuclear Research, Russian Academy of Science, 117312 Moscow, Russia}
\date{\today}

\begin{abstract}
The symmetry energy and the isoscaling properties of the fragments produced in the multifragmentation of 
$^{40}$Ar, $^{40}$Ca + $^{58}$Fe, $^{58}$Ni reactions at 25 - 53 MeV/nucleon were investigated within the 
framework of statistical multifragmentation model. The isoscaling parameters $\alpha$, from the primary (hot) 
and secondary (cold) fragment yield distributions, were studied as a function of excitation energy, isospin 
(neutron-to-proton asymmetry) and fragment symmetry energy. It is observed that the isoscaling parameter $\alpha$ 
decreases with increasing excitation energy and decreasing symmetry energy. The parameter $\alpha$ is also observed 
to increase with increasing difference in the isospin of the fragmenting system. The sequential decay of the 
primary fragments into secondary fragments, when studied as a function of excitation energy and isospin of the 
fragmenting system, show very little influence on the isoscaling parameter. The symmetry energy however, has a strong
influence on the isospin properties of the hot fragments. The experimentally observed scaling parameters
can be explained by symmetry energy that is significantly lower than that for the ground state nuclei near saturation
density. The results indicate that the properties of hot nuclei at excitation energies, densities 
and isospin away from the normal ground state nuclei could be significantly different.
\end{abstract}


\maketitle

\section{Introduction}

Recently the possibility of extracting information on the symmetry energy and the isospin (neutron-to-proton
ratio) of the fragments in a multifragmentation reaction has gained tremendous importance \cite{SHE05,
FEV05,HEN05,BOT04,BOT02}. Such information is of importance for understanding some of the key problems in 
astrophysics \cite{LAT91,LEE96,PET95,LAT00,HIX03,LAT01,BOT04,BOT05}, and various aspects of 
nuclear physics such as the structure of exotic nuclei (the binding energy and rms radii)
\cite{BRO00,HORO01,FUR02,OYA98} and the dynamics of heavy ion collisions \cite{ONO03,CHE05,SHE04,TSA04,BAL98,BAR05,BAR03,LIU04}. 
Traditionally, the symmetry energy of nuclei has been extracted by fitting the binding energy in their 
ground state with various versions of the liquid drop mass formula \cite{LUN03}. The properties of nuclear 
matter are then determined by theoretically extrapolating the nuclear models designed to study the structure 
of real nuclei. However, real nuclei are cold, nearly symmetric ($N \approx Z$) and found at equilibrium density. It is
not known how the symmetry energy behaves at temperatures, isospin (neutron-to-proton ratio) and 
densities away from the normal nuclear matter. Theoretical many-body calculations \cite{DIE03,ZUO99,BRA85,PEA00} 
and those from the empirical liquid drop mass formula \cite{MYE66,POM03} predict symmetry 
energy near normal nuclear density ($\approx$ 0.17 fm$^{-3}$) and temperature ($T$ $\approx$ 0 MeV), to be 
around 28 - 32 MeV. 
\par
In a multifragmentation reaction, an excited nucleus expands to a sub-nuclear density and disintegrates into 
various light and heavy fragments \cite{BOR01,BOW91,AGO96,BEA00}. The fragments are highly excited and 
neutron-rich; their yields depend on the available free energy which in turn depends on the strength 
of the symmetry energy and the extent to which the fragments expand. By studying the isotopic yield 
distribution of these fragments, one can extract important information about the symmetry energy and the 
properties of the fragments at densities, excitation energies and isospin away from those of ground state 
nuclei. 
\par
Experimentally, the determination of fragment isotopic yield distribution is not straight forward. It 
is influenced by the complex de-excitation of the hot (primary) fragments into observed cold (secondary)
fragments. Theoretical calculations of the secondary de-excitations require accurate 
accounting of the feeding from the particle unstable states and are subject to uncertainties in 
the levels that can be excited, and the structure effects that govern their decay \cite{TAN03,XI96,XI99}. 
\par
From various statistical model approaches \cite{TSAN01,BOT02}, it has been shown that the ratio of primary fragment 
yield
for a given isotope or isotone produced in two different reactions with similar temperature, exhibit an 
exponential dependence on proton and neutron number, an observation known as isoscaling. The dependence has 
been interpreted in terms of a scaling parameter that is related to the symmetry energy 
of the primary fragment binding energy. The scaling parameter has been shown to be independent of the complex nature
of the secondary de-excitation and is thus a robust observable for studying the fragment isotopic yield distribution. 
\par
In a recent work \cite{SHE05}, it was shown that the symmetry energy of the primary fragments deduced from the reduced
neutron density is significantly lower than that for the normal nuclei at saturation density. Le Fevre {\it {et al.}}
\cite{FEV05}, in their recent work studied the fragmentation of excited target residues following collisions of $^{12}$C
on $^{112,124}$Sn, at incident energies of 300 and 600 MeV/nucleon. They observed that the symmetry energy
co-efficient deduced from the data are near 25 MeV for peripheral collisions and lower than 15 MeV for the
central collisions. Henzlova {\it {et al.}} \cite{HEN05}, studied the fragments produced in the multifragmentation of
$^{136, 124}$Xe projectiles in mid-peripheral collisions with a lead target at 1 GeV/nucleon. They used
both, the $<N>/Z$ ratio and the isoscaling of the fragments and found that the experimentally determined
value of the scaling parameter can be reproduced within the statistical model framework by lowering the symmetry energy 
to as low as 11 - 12 MeV. The $<N>/Z$ ratios of the fragments on the other hand, could be reproduced with symmetry 
energy co-efficient as low as 14 - 15 MeV. 
\par
In this work, we study the primary fragment yield distribution in a number of reactions at various 
excitation energies and isospins using the isoscaling approach and the equilibrium statistical multifragmentation 
model. It is observed that the isoscaling parameter $\alpha$ for the hot fragments decreases with 
increasing excitation energy and decreasing symmetry energy. The alpha values also increase with increasing
difference in the isospin (neutron-to-proton asymmetry) of the fragmenting system. A similar behavior is also
observed for the cold secondary fragments. The secondary de-excitation is found to have very little 
influence on the isoscaling parameter at lower excitation energies and isospin of the fragmenting system.
The symmetry energy however, strongly influence the properties of the hot fragments. The experimentally 
determined isoscaling parameters can be explained by symmetry energy that is significantly lower than that 
for the normal (cold) nuclei at saturation density, indicating that the properties of nuclei at high
excitation energy, isospin and reduced density are sensitive to the symmetry energy. 
\par
The paper is organized as follows. In sec. II, we describe the experiment in detail. The experimental results 
are presented and discussed in Sec. III. Section IV contains a brief description of the statistical 
multifragmentation model used in the analysis. Section V contains a comparison between the experimental 
data and the statistical multifragmentation model. Finally, a summary and conclusion is given in section VI.

\section{Experiment}
\subsection{Experimental Setup}
%Fig 1
    \begin{figure}
    \includegraphics[width=0.52\textwidth]{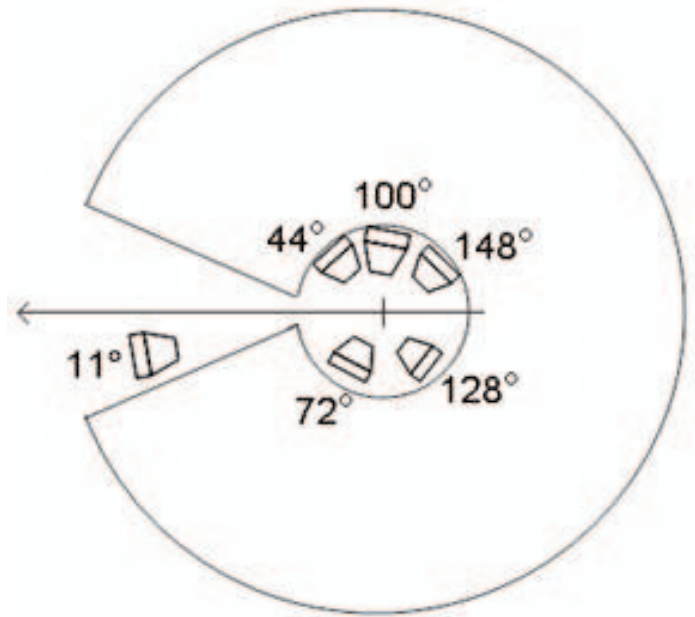}
    \caption{Schematic diagram of the experimental setup showing the placement of the telescopes inside
    the scattering chamber of the neutron ball.}
    \end{figure}
The experiments were carried out at the Cyclotron Institute of Texas A$\&$M University (TAMU) using the K500 
Superconducting Cyclotron where targets of $^{58}$Fe (2.3 mg/cm$^{2}$)  and $^{58}$Ni (1.75 mg/cm$^{2}$) were 
bombarded with beams of $^{40}$Ar and $^{40}$Ca at 25, 33, 45 and 53 MeV/nucleon \cite{JOH97}. The various combinations of 
target and projectile nuclei allowed for a range of $N/Z$ (neutron-to-proton ratio) (1.04 - 1.23) of the system 
to be studied, while keeping the total mass constant ($A$ = 98). The beams were fully stripped by allowing them 
to pass through a thin aluminum foil before being hit at the center of the target inside the TAMU 4$\pi$ neutron 
ball \cite{SCH95}.  Light charged particles ($Z$ $\leq$ 2) and intermediate mass fragments ($Z$ $>$ 2) were 
detected using six discrete telescopes placed inside the scattering chamber of the neutron ball at angles  of 
10$^{\circ}$, 44$^{\circ}$, 72$^{\circ}$, 100$^{\circ}$, 128$^{\circ}$ and 148$^{\circ}$. A schematic
diagram of the setup showing placement of the telescopes within the scattering
chamber of the neutron ball is shown in figure. 1. Each telescope consisted of a gas ionization 
chamber (IC) followed by a pair of silicon detectors (Si-Si) and a CsI scintillator detector, providing three 
distinct detector pairs 
(IC-Si, Si-Si, and Si-CsI) for fragment identification. The ionization chamber was of axial field design and was 
operated with CF$_{4}$ gas 
at a pressure of 50 Torr. The gaseous medium was 6 cm thick with a typical threshold of $\sim$ 0.5 MeV/nucleon for 
intermediate mass 
fragments. The silicon detectors had an active area of 5 cm $\times$ 5 cm and were each subdivided into four quadrants. 
The first and second 
silicon detectors in the stack were 0.14 mm and 1 mm thick, respectively. The dynamical energy range of the silicon 
pair was  $\sim$ 16 - 50 MeV 
for $^{4}$He and  $\sim$ 90 - 270 MeV for $^{12}$C. The CsI scintillator crystals that followed the silicon detector 
pair were 2.54 cm in 
thickness and were read out by photodiodes.  Good elemental ($Z$) identification was achieved for fragments 
that punched through the IC detector 
and stopped in the first silicon detector. Fragments measured in the Si-Si detector pair also had good
isotopic separation. The trigger for the data acquisition was generated by requiring  a valid hit in one of 
the silicon detectors.

\subsection{Calibration}
The calibration of the IC-Si detectors were carried out using the standard alpha sources and by operating the IC 
at various gas pressures. The 
Si-Si detectors were calibrated by measuring the energy deposited by the alpha particles in the thin silicon and 
the punch-through energies  of 
different isotopes in the thick silicon. The Si-CsI detectors were calibrated by selecting points along the 
different light charged isotopes and 
determining the energy deposited in the CSI crystal from the energy loss in the calibrated Si detector. 

\subsection{Neutron Multiplicity and Event Characterization} 
Neutrons were measured with the 4$\pi$ neutron ball that surrounded the detector assembly. 
The neutron ball consisted of eleven scintillator tanks segmented in its median plane and 
surrounding the vacuum chamber. The upper and the lower tank were 1.5 m diameter hemispheres. Nine 
wedge-shaped detectors were sandwiched between the hemispheres. All the wedges subtended 40$^{\circ}$
in the horizontal plane. The neutron ball were filled with a 
pseudocumene based liquid scintillator mixed with 0.3 weight percent of Gd salt (Gd 2-ethyl hexanoate). Scintillations 
from thermal neutrons captured by Gd were detected by twenty 5-in phototubes : five in each
hemisphere, one on each of the identical 40$^{\circ}$ wedges and two on the forward edges. The
efficiency with which the neutrons could be detected is 
about 83$\%$, as measured with a $^{252}$Cf source.
\par
The detected neutrons were used to differentiate between the central and peripheral collisions. To understand the effectiveness 
of neutron multiplicity as a centrality trigger, simulations were carried out using a hybrid BUU-GEMINI
calculations at various impact parameters for the $^{40}$Ca + $^{58}$Fe reaction at 33 MeV/nucleon. The simulated neutron 
multiplicity distribution was compared with the experimentally measured distribution. The multiplicity of neutron 
for the impact parameter {\it {b}} = 0 collisions was found to be higher than the {\it {b}} = 5 collision. By gating on the 10$\%$
highest neutron multiplicity events, one could clearly discriminate against the peripheral events. 
\par
To determine the contributions from noncentral impact parameter collisions,  neutrons emitted in coincidence with fragments at 44$^{\circ}$ and 
152$^{\circ}$ were calculated at {\it {b}} = 0 fm and {\it {b}} = 5 fm. The number of events were adjusted 
for geometrical cross sectional differences. A ratio was made between the number of events with a neutron multiplicity of at least six, 
calculated at {\it {b}} = 0 fm, and the number of events with the same neutron multiplicity at {\it {b}} =
5 fm. The ratios were observed to 
%Fig 2
    \begin{figure}
    \includegraphics[width=0.5\textwidth,height=0.5\textheight]{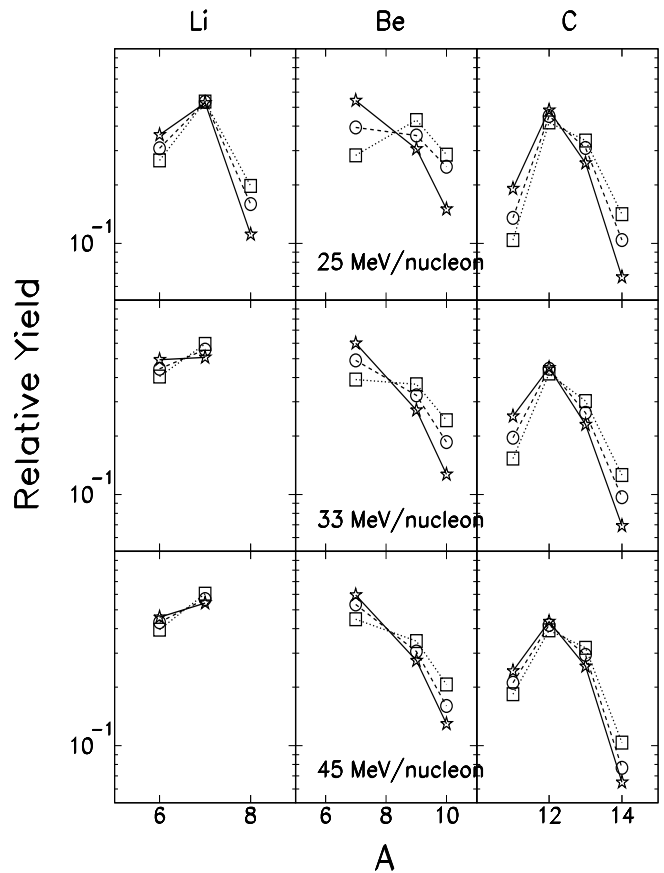}
    \caption{Relative isotopic yield distribution for the Lithium (left), Berillium(center) and Carbon
    elements in $^{40}$Ca + $^{58}$Ni (stars and solid lines), $^{40}$Ar + $^{58}$Ni (circles and dashed
    lines) and $^{40}$Ar + $^{58}$Fe (squares and dotted lines) reactions at various beam energies.}
    \end{figure}
be 19.0 and 11.1 at 44$^{\circ}$ and 1.3 and 2.2 at 152$^{\circ}$ for 33 and 45 MeV/nucleon respectively. At intermediate angles, high neutron 
multiplicities were observed to be outside the region in which {\it {b}} = 5 fm contributes significantly.
At backward angles the collisions at {\it {b}} = 5 fm made a larger contribution to the neutron multiplicity.
\par
In addition to the neutron
multiplicity distribution, the charge distribution of the fragments was also used to investigate the 
contributions from central and mid-impact parameter collisions. The {\it {b}} = 5 collisions produced essentially no fragments with  charge 
greater than three in the 44$^{\circ}$ telescope.
\par
In an earlier work \cite{JOH97}, some analysis of the fragment kinetic energy and charge distributions were
presented. It was shown that at a laboratory angle of 44$^{\circ}$ the kinetic energy and the charge
distributions are well reproduced by the statistical model calculation. Using a moving source analysis of
the fragment energy spectra, it was also shown that the fragments emitted at backward angles originate
from a target-like source, while those emitted at 44$^{\circ}$ originate primarily from a composite
source. In this work, we will concentrate exclusively on data from the laboratory angle
of 44$^{\circ}$ to study the symmetry energy and the isoscaling properties of the fragments
produced. The choice of this angle 
enables one to select events which are predominantly central and undergo bulk multifragmentation. 
The contributions to the intermediate mass fragments from the projectile-like and target-like sources 
can therefore be assumed to be minimal and the use of equilibrium statistical model appropriately justified.

\section{Experimental Results}

\subsection{Relative Fragment Yield}
The experimentally measured relative isotopic yield distributions for the Lithium (left),
Berillium (center) and Carbon (right) elements, in $^{40}$Ca + $^{58}$Ni (star symbols),
$^{40}$Ar + $^{58}$Ni (circle symbols) and $^{40}$Ar + $^{58}$Fe (square symbols) reactions, are shown in
figure 2 for the beam energies of 25, 33 and 45 MeV/nucleon. The distributions for each element show higher 
fragment yield for the neutron rich isotopes in $^{40}$Ar + $^{58}$Fe reaction (squares), which has the 
largest neutron-to-proton ratio ($N/Z$),  in comparison to the $^{40}$Ca + $^{58}$Ni reaction (stars), which 
has the smallest neutron-to-proton ratio. The yields for the reaction, $^{40}$Ar + $^{58}$Fe (circles), which has an 
intermediate value of the neutron-to-proton ratio, are in between those of the other two reactions. The 
figure thus shows the isospin dependence of the composite system on the properties of the
fragments produced in the multifragmentation reaction. One 
also observes that the relative difference in the yield distribution between the three reactions 
decreases with increasing beam energy. This is due to the secondary de-excitation of the primary 
fragments, a process that becomes important for systems with increasing neutron-to-proton ratio and excitation 
energy. In the following subsections, we will utilize the experimentally determined isotopic yield
distributions to establish the isoscaling properties of the produced fragments before comparing
them with the statistical multifragmentation model in section V.

%Fig 3
    \begin{figure}
    \includegraphics[width=0.47\textwidth,height=0.51\textheight]{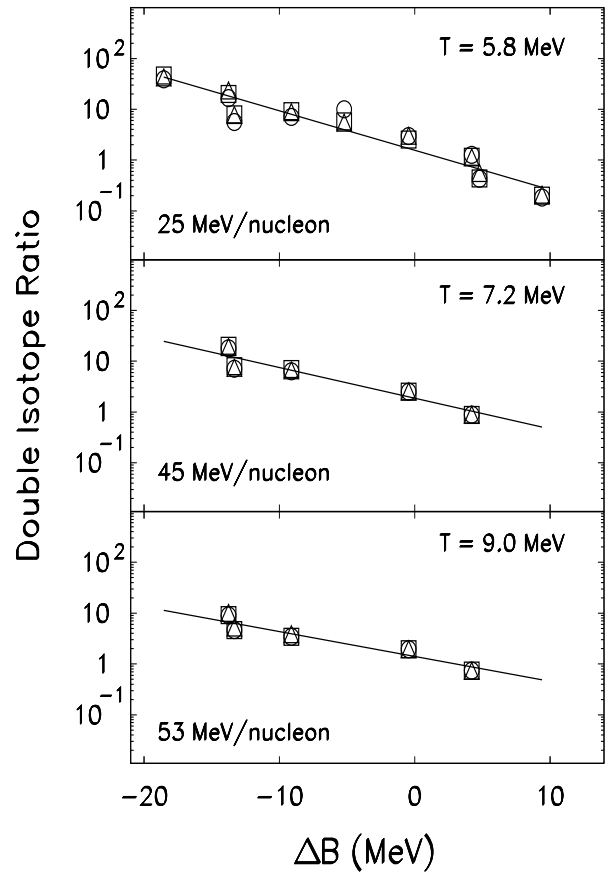}
    \caption{Double isotope ratio as a function of the difference in binding energy for various beam
     energies. The circle symbols correspond to $^{40}$Ca + $^{58}$Ni reaction, square symbols to 
     $^{40}$Ar + $^{58}$Ni reaction, and triangle symbols to $^{40}$Ar + $^{58}$Fe reaction. The solid
     lines are the best fit to the data.}
    \end{figure}

%Fig 4
    \begin{figure}
    \includegraphics[width=0.47\textwidth,height=0.53\textheight]{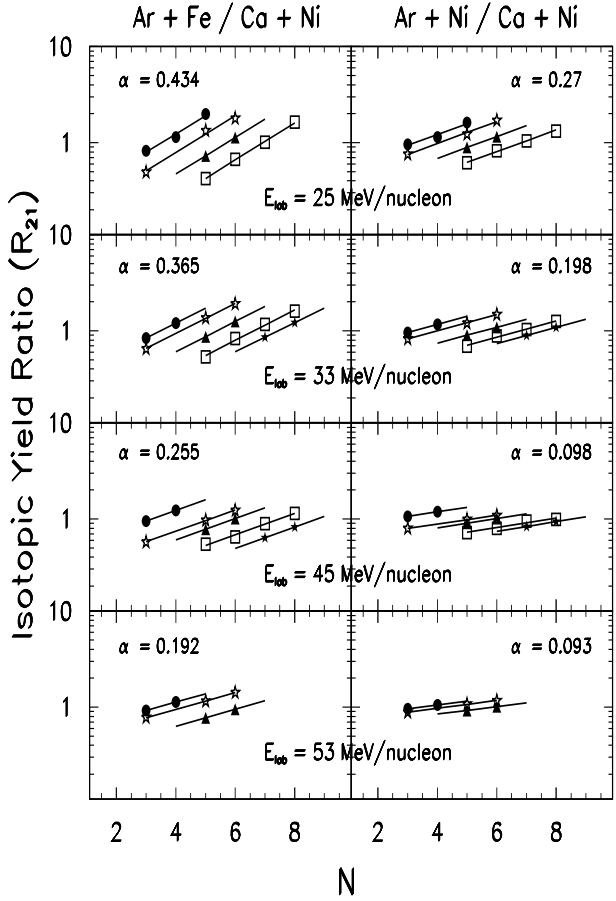}
    \caption{Experimental isotopic yield ratios of the fragments as a function of neutron number $N$, for various beam energies. The left
    column correspond to $^{40}$Ar + $^{58}$Fe and $^{40}$Ca + $^{58}$Ni pair of reactions. The right
    column correspond to $^{40}$Ar + $^{58}$Ni and $^{40}$Ca + $^{58}$Ni pair of reactions. The different
    symbols correspond to $Z$ = 3 (circles), $Z$ = 4 (open stars), $Z$ = 5 (triangles), $Z$ = 6 (squares) and
    $Z$ = 7 (filled stars) elements. The lines are the exponential fits to the data as explained in the text.}
    \end{figure}

%Fig 5
    \begin{figure}
    \includegraphics[width=0.47\textwidth, height=0.53\textheight]{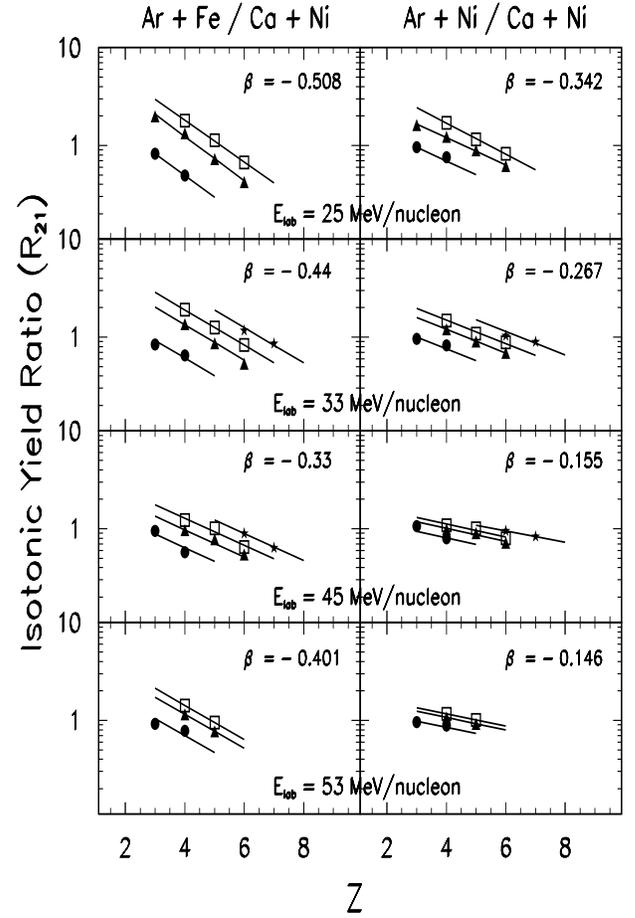}
    \caption{Experimental isotonic yield ratios of the fragments as a function of proton number $Z$, for various beam energies. The left
    column correspond to $^{40}$Ar + $^{58}$Fe and $^{40}$Ca + $^{58}$Ni pair of reaction. The right
    column correspond to $^{40}$Ar + $^{58}$Ni and $^{40}$Ca + $^{58}$Ni pair of reaction. The different
    symbols correspond to $N$ = 3 (circles), $N$ = 5 (triangles), $N$ = 6 (squares) and $N$ = 7
    (stars) elements. The lines are the exponential fits to the data as explained in the text.  }
    \end{figure}

%Fig 6
    \begin{figure}
    \includegraphics[width=0.49\textwidth,height=0.43\textheight ]{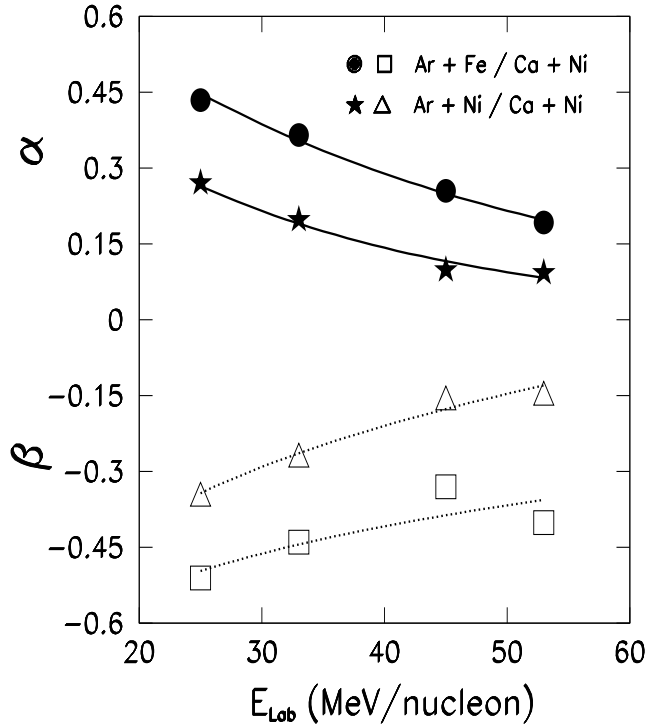}
    \caption{Isoscaling parameters $\alpha$ (solid symbols) and $\beta$ (open symbols) as a
    function of the beam energy. The solid circles and open squares correspond to $^{40}$Ar + $^{58}$Fe and $^{40}$Ca +
    $^{58}$Ni pair of reactions. The solid stars and open triangles correspond to $^{40}$Ar + $^{58}$Ni and $^{40}$Ca +
    $^{58}$Ni pair of reactions. The lines are the exponential fits to the data. The error bars are
    of the size of the symbols.}
    \end{figure}

%Fig 7
    \begin{figure}
    \includegraphics[width=0.46\textwidth,height=0.51\textheight]{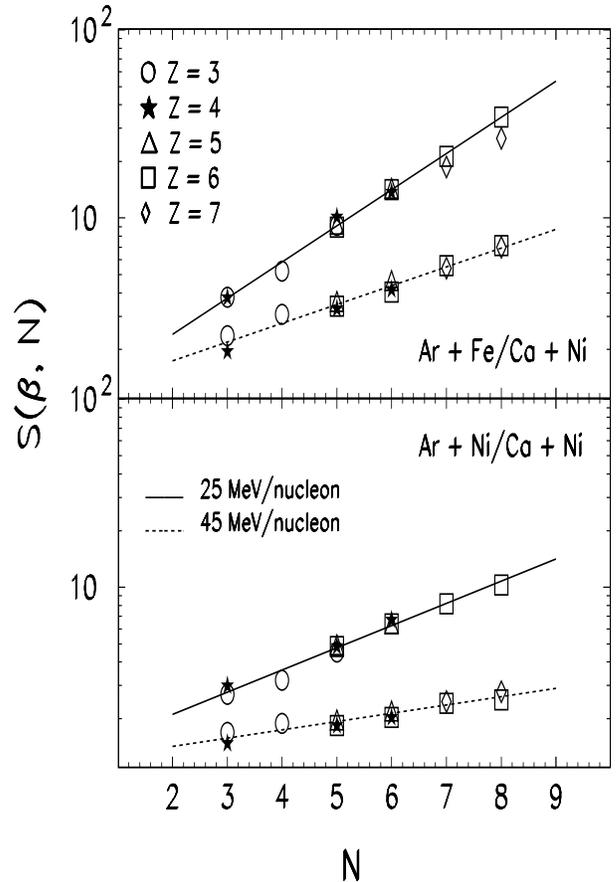}
    \caption{The scaled isotope ratio $S(\beta , N)$, as a function of the neutron number $N$, for
    the 25 and 45 MeV/nucleon beam energies. The top panel is for the $^{40}$Ar +
    $^{58}$Fe/$^{40}$Ca + $^{58}$Ni pair, and the bottom panel is for the $^{40}$Ar +
    $^{58}$Ni/$^{40}$Ca + $^{58}$Ni pair of reactions. The symbols correspond to $S(\beta , N)$ obtained from
    various elements ($Z$). The lines are the best fits to the data.}
    \end{figure}

%Fig 8
    \begin{figure}
    \includegraphics[width=0.47\textwidth]{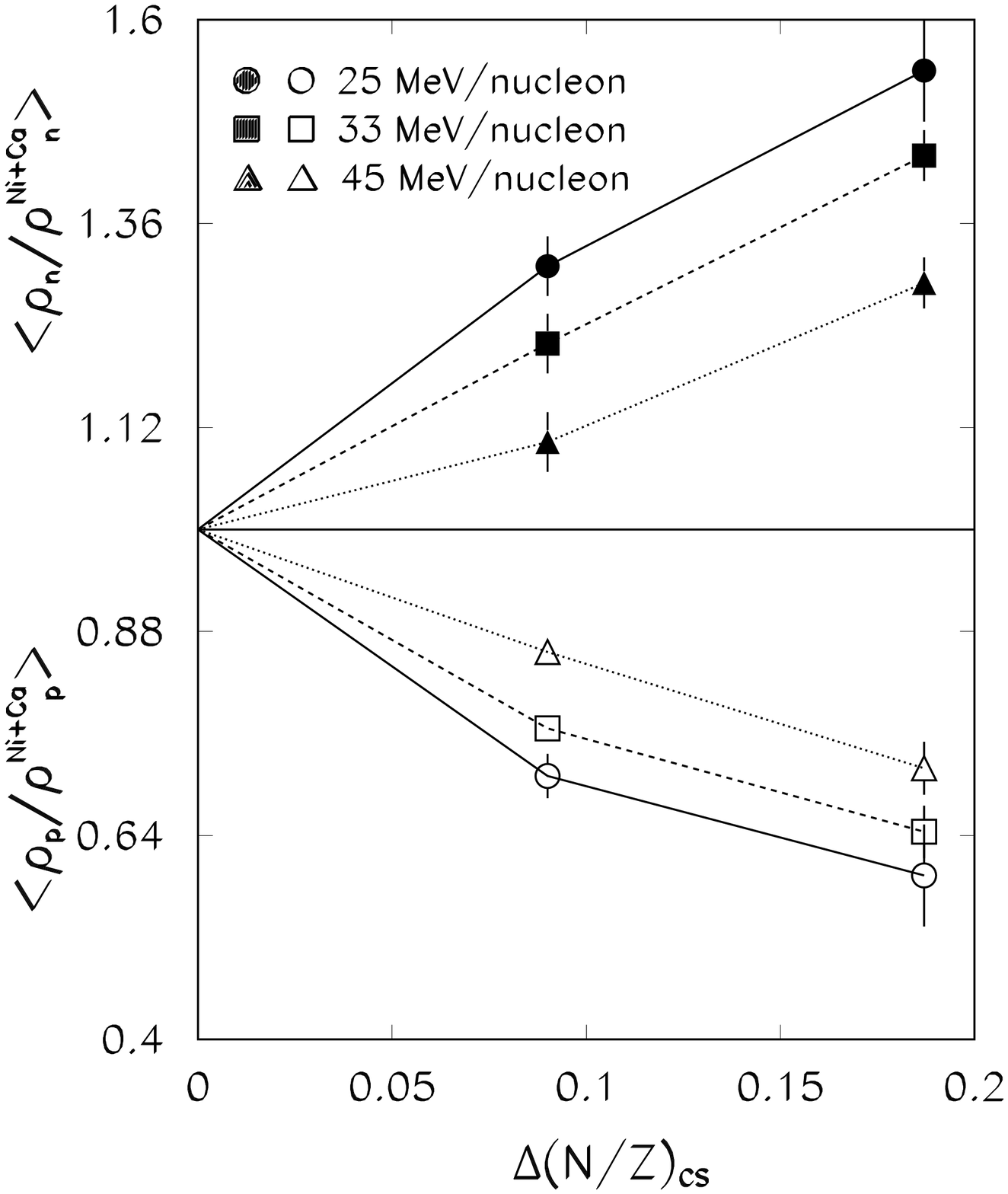}
    \caption{Relative free neutron (solid symbols) and proton (open symbols) densities as a function of the difference in $N/Z$ of the
    systems for the beam energies of 25, 33 and 45 MeV/nucleon.}
    \end{figure}

\subsection{Isotopic and Isotonic Scaling}
In a multifragmentation reactions, the ratio of isotope yields in two different systems, 1
and 2, $R_{21}(N,Z) = Y_{2}(N,Z)/Y_{1}(N,Z)$, has been shown to obey an exponential dependence on the
neutron number ($N$) and the proton number ($Z$) of the isotopes, an observation known as isoscaling
\cite{BOT02, TSAN01, TSANG01, RAD05}. The dependence is characterized by a simple relation,

\begin{equation}
              R_{21}(N,Z) = Y_{2}(N,Z)/Y_{1}(N,Z) = C exp({\alpha N + \beta Z})
\end{equation}

where, $Y_{2}$ and $Y_{1}$ are the yields from the neutron-rich and neutron-deficient systems,
respectively. $C$, is an overall normalization factor and $\alpha$ and $\beta$ are the parameters
characterizing the isoscaling behavior.
\par
The necessary condition for observing isoscaling in multifragmentation reaction is the near equality of 
temperature for systems chosen in the above scaling relation \cite{TSA01}. For the present work, this condition 
was tested by determining the temperature using the double isotope ratio method of Albergo {\it {et al.}} \cite{ALB85}.  
The Albergo method relates the apparent temperature $T$ of the system at the freeze-out to the double
isotope ratio, $R = [Y(N,Z_{1})/Y(N + 1, Z_{1})]/[Y(N,Z_{2})/Y(N + 1, Z_{2})]$, through a relation

\begin{equation}
      T = \frac{\Delta B}{ln(a R)}
\end{equation}

where $a$ is a factor that depends on the statistical weights of the ground state nuclear spins, and $\Delta
B = [B(N,Z_{1}) - B(N + 1, Z_{1})] -  [B(N,Z_{2}) - B(N + 1, Z_{2})]$ is the difference in the binding
energy. The method was applied to all the three systems and beam energies studied. 
\par
Figure 3 shows the double isotope ratios obtained using the various combinations of isotopes as a 
function of the difference in the binding energy. The top panel correspond to those determined for the beam energies
of 25 MeV/nucleon, the center for the 45 MeV/nucleon and the bottom for the 53 MeV/nucleon. The different 
symbols correspond to the three different reactions studied. The 
isotope ratios for the three reactions at each beam energy overlap nicely indicating formation of composite systems with similar
temperature/excitation energy. The slopes of the exponential fits to the data (shown by solid line), which correspond to
the apparent temperatures of the systems, show a gradual decrease with increasing beam energy indicating an increase in the 
excitation energy/temperature of the system.
\par
Having satisfied the necessary condition for isoscaling, 
the isotopic yield ratio as a function of neutron number $N$, for the beam energies of 25, 33, 45 and 53
MeV/nucleon is plotted in Fig 4. The left column shows the ratio for the $^{40}$Ar + $^{58}$Fe and $^{40}$Ca + $^{58}$Ni pair
of reaction and the right column shows the ratio for the $^{40}$Ar + $^{58}$Ni and $^{40}$Ca + $^{58}$Ni pair
of reaction. One observes that the ratio for each element lies along a straight line in the logarithmic plot and
align with the neighboring element quite well. This feature is observed for all the beam energies
and both pairs of reaction studied. One observes that the alignment of the data points varies with
beam energies as well as the pairs of reaction. 
To have a quantitative estimate of this variation, the ratio for each elements ($Z$) were simultaneously fitted using an exponential 
relation (shown by the solid lines) to obtain the slope parameter $\alpha$. The values of the parameters are shown at the top of each 
panel in the figure. The value of the slope parameter $\alpha$ is larger for the 
$^{40}$Ar + $^{58}$Fe and $^{40}$Ca + $^{58}$Ni reactions, which has a larger difference in the $N/Z$
of the systems in the pair, compared to the $^{40}$Ar + $^{58}$Ni and $^{40}$Ca + $^{58}$Ni reactions, which has 
a smaller difference in the corresponding $N/Z$. The $\alpha$ value furthermore 
decreases with increasing beam energy. Figure 5. shows the isotonic yield ratio as a function of atomic
number $Z$, for the same beam energies and pairs of systems as shown in Figure 4. Once again, one observes
the ratios for each isobar to align nicely with each other at all beam energies. The scaling parameters
$\beta$ in this case, shows an increase with increasing beam energy. The values of the $\beta$ parameter
are larger for the $^{40}$Ar + $^{58}$Ni and $^{40}$Ca + $^{58}$Ni reaction pair compared to the 
$^{40}$Ar + $^{58}$Fe and $^{40}$Ca + $^{58}$Ni reaction pair. Figure 6 shows a relative comparison
of how the $\alpha$ and $\beta$ parameters evolve as a function of beam energy and the isospin of the system.
\par
The temperature/excitation energy 
dependence of the isoscaling properties can  be further studied by constructing the scaling factor,
$S(\beta , N)$ = $R_{21}(N,Z)e^{- \beta Z}$. The scaling factor is known to be a robust feature over a large range of data, 
from deep inelastic heavy ion reactions at lowest energies, through evaporation reactions 
induced by light-ion and heavy-ion projectiles, to high energy heavy ion reactions characterized by intermediate 
mass fragments and multifragmentation \cite{TSANG01}. With a single value for the $\beta$ parameter, all the isotopes 
should fall along a single line in a plot of $S(\beta , N)$ vs $N$. This is shown in Fig. 7, where the
$S(\beta , N)$ 
from two different reaction pairs are plotted as a function of neutron number $N$ 
for beam energies of 25 MeV/nucleon and 45 MeV/nucleon. The parameter $\beta$, was taken  from the fit 
to the isotonic yield ratio shown in figure 5. As shown in figure 7, the values of $S(\beta , N)$ obtained from various
elements ($Z$) cluster and scale along a single line. The figure shows a significant difference in the scaling 
for the two beam energies, indicating the influence of temperature on the isotopic yields of the 
light clusters. The present observation alternatively demonstrates the role played by the temperature in the 
distillation of nuclear matter into a neutron-rich gas and a symmetric liquid phase. We illustrate this further 
in the following section.

\subsection{Isospin fractionation and the reduced nucleon densities}

In Grand-Canonical approach of the multifragmentation process (see e.g. \cite{ALB85,BON95,RAN81,BOT87}), 
the fragment yield with neutron number $N$, and proton number $Z$ (mass number $A = N + Z$), can be
written as

\begin{equation}
 Y(N,Z) \propto V \rho_{n}^{N}\rho_{p}^{Z}Z_{N,Z}(T)A^{3/2}e^{B(N,Z)/T}
\end{equation}

where $V$ is the volume of the system and $\rho_{n}$ ($\propto$ e$^{\mu_{n}/T}$) and $\rho_{p}$ ($\propto$
e$^{\mu_{p}/T}$) are 
the primary `free' neutron and proton densities. The exponents $\mu_{n}$ and $\mu_{p}$ are the neutron and the
proton chemical potentials, and $Z_{N,Z}(T)$ is the intrinsic partition function of the excited fragment. The quantity $B(N,Z)$, is 
the ground state binding energy of the fragment and $T$ is the temperature. In the above formula, the effect of Coulomb 
interaction on fragment yield is neglected by introducing $\rho_{n}$ and $\rho_{p}$. The actual isotope
yields then reduce to an approximation appropriate for the thermodynamical limit at high excitation
energy \cite{BON95}. 
As discussed in the introduction, taking the ratios of the fragment yields from two different systems which differ only in their 
isospin $(N/Z)$ content reduces the uncertainties in the quantities shown in Eq. 3. The isotopic yield distribution of the fragments in 
terms of relative reduced neutron density can then be written as,  

\begin{equation}  
   \frac{ Y(N + k, Z)/Y^{Ca + Ni}(N + k, Z)} { Y(N, Z)/Y^{Ca + Ni}(N, Z)  } = \bigg(\frac{\rho_n}{\rho_{n}^{Ca + Ni}} \bigg )^{k} ,
\end{equation}

where $k$, corresponds to various isotopes of an element that can be used to determine the double ratio.
The quantity $Y^{Ca + Ni}$ is the yield for the $^{40}$Ca + $^{58}$Ni reaction with respect to which all the 
ratios are taken in this work. A similar expression for the relative reduced proton density from the isotonic yield ratios can 
also be written as

\begin{equation}  
   \frac{ Y(N, Z + k)/Y^{Ni + Ni}(N, Z + k)} { Y(N, Z)/Y^{Ca + Ni}(N, Z)  } = \bigg(\frac{\rho_p}{\rho_{p}^{Ca + Ni}} \bigg )^{k} ,
\end{equation}

In the statistical limit for a dilute non-interacting gas, the relative nucleon densities are related to
the isoscaling parameters $\alpha$ and $\beta$, through a relation, 
$\rho_{n}^{Ar + Fe}/\rho_{n}^{Ca + Ni}$ = $e^{\alpha}$ and $\rho_{p}^{Ar + Fe}/\rho_{p}^{Ca + Ni}$ =
$e^{\beta}$, where $\alpha = \Delta \mu_n/T$ and $\beta = \Delta \mu_p/T$, with $\Delta \mu_n$ and
$\Delta \mu_p$ being the difference in neutron and proton chemical potentials.
\par
In figure 8, we show the experimentally obtained relative reduced neutron and proton density as a function of the difference in the
$N/Z$ of the systems for the Ar + Ni/Ca + Ni and Ar + Fe/Ca + Ni pairs of reactions. All densities shown are relative to those of Ca + Ni reaction. 
The circle symbols correspond to the 25 MeV/nucleon, squares to the 33 MeV/nucleon and the triangles to
the 45 MeV/nucleon beam energies.
The $\alpha$ and the $\beta$ values were taken from the fit to the isotopic and isotonic yield
ratios of figures 4 and 5. The figure shows  a steady decrease in the reduced neutron density and an 
increase in the proton density with increasing beam energies. The effect is stronger for the $^{40}$Ar +
$^{58}$Fe and $^{40}$Ca + $^{58}$Ni reaction pair which has the highest difference in $N/Z$.
\par
An important feature of the data shown in figure 8, is the decrease in the relative neutron-proton
asymmetry ($\rho_{n}/\rho_{n}^{Ca + Ni} - \rho_{p}/\rho_{p}^{Ca + Ni}$) with increasing beam energy. The asymmetry 
is found to decrease from $\sim$ 1.0 at 25 MeV/nucleon to $\sim$ 0.6 at 45 MeV/nucleon for the Ar + Fe/Ca + Ni pair of reaction. 
This is consistent with the prediction of phase coexistence of two conserved charges with isospin dependent interaction of lattice gas 
calculation \cite{BAL01,CHO99}. The large asymmetry or the strong enrichment of the gas phase can be understood in terms of an isospin 
dependent interaction between the protons and the 
neutrons. In the absence of an isospin dependent interaction, the fragments are populated close to the stability line and 
have similar $N/Z$ ratio's as the initial source. With the inclusion of isospin dependence, the heavy
fragments are strongly favored  from the energetic point of view to be populated closer to the 
bottom of the stability valley. Only a few protons are left to be shared between 
the light fragments leading to the observed enhancement of neutrons in the gas phase. With the increase in temperature, the fragment 
distribution is centered closer to the $N = Z$ line resulting in a decrease in the asymmetry of the gas phase. 
\par
The observed decrease in the relative neutron and proton densities with increasing beam energy may thus
be attributed to the decrease in the sensitivity of the isospin effect with increasing temperature. The
scaling parameter is also sensitive to the breakup density, and the observed decrease in the
neutron-proton density could result from a decrease in the break-up density, though it is expected
to be less sensitive. 
    
\section{Statistical Multifragmentation Model}

Statistical models \cite{BON95,RAN81,BOT87,GRO90,GRO82,ZHA87} are widely used
for describing multifragmentation 
reactions \cite{BOT95,SCH01,AGO96,BEL02,AVD02,POC95,AGO99}. They are based on the assumption of statistical equilibrium 
at a low density freeze-out stage. In the Statistical Multifragmentation Model (SMM) \cite{BON95, BOT01},
all break-up channels composed of nucleons and excited fragments are taken into account and considered as 
partitions.  During each partition, the conservation of mass, charge, energy and angular momentum is taken 
into account, and the partitions are sampled uniformly in the phase space according to their statistical weights using 
the Monte Carlo sampling. In the present calculations the 
Coulomb interaction between the fragments is treated in the Wigner-Seitz approximation. Light fragments 
with mass number $A$ $\leq$ 4 are considered as elementary particles with only translational
degrees of freedom (``nuclear gas"). Fragments with 
$A$ $>$ 4 are treated as heated nuclear liquid drops, and their individual free energies $F_{A,Z}$ are parametrized as a sum of the volume, surface, 
Coulomb, and symmetry energy,

\begin{equation}
            F_{A,Z} = F^{V}_{A,Z} + F^{S}_{A,Z} + E^{C}_{A,Z} + E^{sym}_{A,Z}
\end{equation}

where $F^{V}_{A,Z} = (-W_{o} - T^{2}/\epsilon_{o})A,$ with parameter $\epsilon_{o}$ related to the level
density and $W_{o}$ = 16 MeV being the binding energy of infinite nuclear matter. $F^{S}_{A,Z} = B_{o}A^{2/3}[(T^{2}_{c} -
T^{2})/(T^{2}_{c} + T^{2})]^{5/4}$, with $B_{o}$ = 18 MeV being the surface co-efficient and $T_{c}$ = 18 MeV being the 
critical temperature of infinite nuclear matter. $E^{C}_{A,Z} = c Z^{2}/A^{1/3}$, where $c = (3/5)(e^{2}/r_{o})[1 - (\rho/\rho_{o})^{1/3}]$, 
is the Coulomb parameter obtained in the Wigner-Seitz approximation with charge unit $e$, and $r_{o}$ = 1.17 fm. 
$E^{sym}_{A,Z} = \gamma (A - 2Z)^{2}/A$, where $\gamma$ = 25 MeV is the symmetry energy co-efficient. These 
parameters are those adopted from the Bethe-Weizsacker mass formula and correspond to the assumption of isolated fragments with normal density in the freeze-out 
configuration. The value of the symmetry energy co-efficient $\gamma$ is taken from the fit to the binding energies of 
isolated cold nuclei in their ground states. In a multifragmentation process the  primary fragments
are not only excited but also expanded. The fragments continue to interact in-medium with each other inside the freeze-out volume 
and modify their parameters. 
By comparing the experimentally determined fragment yield distribution with the SMM calculation, the 
parameters of hot nuclei under multifragmentation conditions, including the symmetry energy, can
be extracted. In the following, it will be shown how this information can be obtained from the isoscaling phenomena.
    
\subsection{Isoscaling and Symmetry Energy Coefficient}

Isotopic scaling or isoscaling arise naturally in statistical equilibrium models of multifragmentation. 
In these models the difference in the chemical potential of systems with different $N/Z$ is directly 
related to the scaling parameter $\alpha$. It has been shown that the isoscaling parameter $\alpha$ is 
proportional to the symmetry energy part of the fragment binding energy through a relation,

\begin{equation}
   \alpha = \frac{4 \gamma}{T} \bigg (\frac{Z_{1}^{2}}{A_{1}^{2}} - \frac{Z_{2}^{2}}{A_{2}^{2}} \bigg )
\end{equation}

where $Z_{1}$, $A_{1}$ and $Z_{2}$, $A_{2}$ are the charge and the mass numbers of the fragmenting
systems, $T$ is the temperature of the system and $\gamma$, the symmetry energy co-efficient \cite{BOT02}.

\subsection{Secondary de-excitation of the fragments}

The above formula in the statistical model approach is valid at the freeze-out stage where the primary hot
fragments are formed at reduced density. In order to extract information on symmetry  energy $\gamma$,
from the observed cold 
secondary fragments, one has to take into account the process of secondary de-excitation. 
In SMM, the secondary de-excitation of large fragments with $A$ $>$ 16 is described by Weisskopf-type
evaporation and Bohr-Wheeler-type fission models \cite{BON95,BOT87}. The decay of smaller fragments is treated with the
Fermi-breakup model. All ground and nucleon-stable excited states of light fragments are
taken into account and the population probabilities of these states are calculated according to the
available phase space \cite{BOT87}. 
%Fig 9
    \begin{figure}
    \includegraphics[width=0.47\textwidth,height=0.51\textheight]{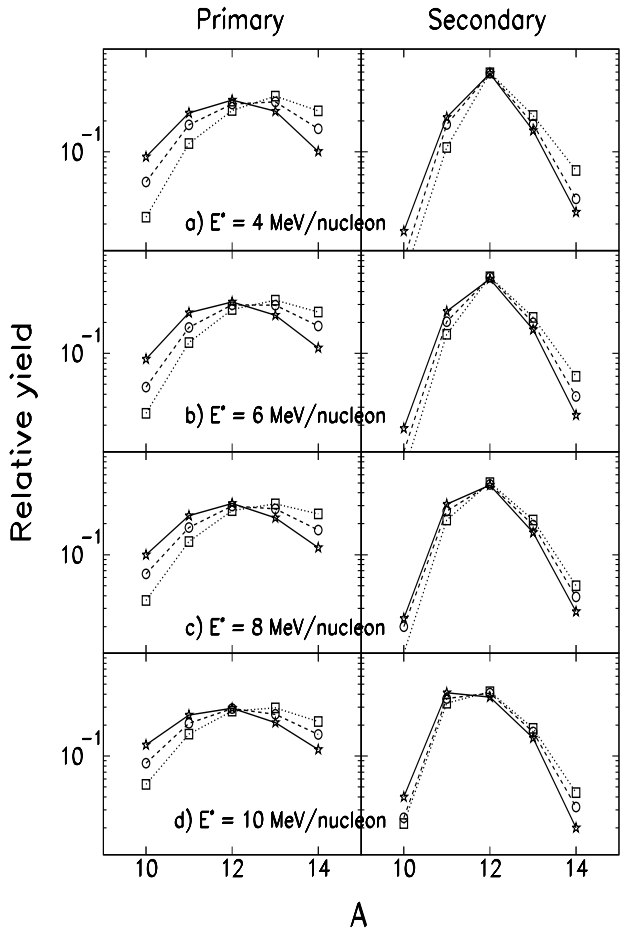}
    \caption{SMM calculated primary (left) and secondary (right) fragment isotope yield
    distributions for the carbon element
    in $^{40}$Ca + $^{58}$Ni (stars and solid lines), $^{40}$Ar + $^{58}$Ni (circles and dashed
    lines) and $^{40}$Ar + $^{58}$Fe (squares and dotted lines) reactions at various excitation
    energies. The calculations are for $\gamma$ = 25 MeV.}
    \end{figure}   
%Fig 10
    \begin{figure}
    \includegraphics[width=0.47\textwidth,height=0.51\textheight]{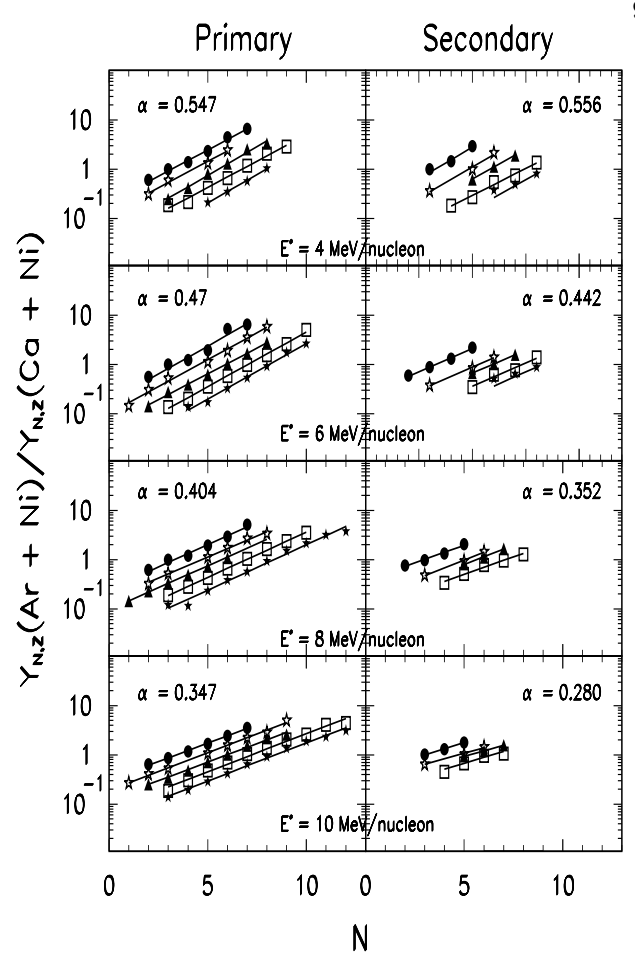}
    \caption{Calculated isotopic yield ratios from the primary (left) and the secondary (right)
    fragment yield distributions for the $^{40}$Ar + $^{58}$Ni and $^{40}$Ca + $^{58}$Ni pairs at various
    excitation energies. The calculations are for $\gamma$ = 25 MeV. The different symbols shown
    correspond to $Z$ = 3 (circles), $Z$ = 4 (open stars), $Z$ = 5
    (triangles), $Z$ = 6 (square) and $Z$ = 7 (solid stars) elements. The lines are the exponential fits 
    to the ratios.}
    \end{figure}    
    
\section{Comparison with the experimental data}
In order to compare the experimentally observed results to the theoretical predictions, we have calculated the primary and 
the secondary fragment isotopic yield distributions using the above described SMM. The calculations were
carried out for the $^{40}$Ca + $^{58}$Ni, $^{40}$Ar + $^{58}$Ni and $^{40}$Ar + $^{58}$Fe reactions at various excitation energies. The 
excitation energy per nucleon of the initial system for the calculation depends strongly on the 
matching condition between the dynamical and statistical stage of the collision. This quantity is presently difficult 
to calculate accurately. A range of values for the excitation energy per nucleon from 
$E^{*}$ = 4 - 10 MeV/nucleon were therefore assumed. The excitation energy corresponding to each beam
energy was also verified by an independent calculation using BUU - GEMINI (see table II in Ref.\cite{JOH97}), and the
systematic calorimetric measurements available in the literature for systems with mass A $\sim$ 100 \cite{CUS93}. 
The mass and the charge of the initial systems were assumed to be those of the initial compound nucleus. 
To check for the possible uncertainty in the source size due to loss of nucleons during
pre-equilibrium emission, the calculation was also performed for sources with 80$\%$ of the total mass.
No significant change in the isospin characteristics under study was observed. The freeze-out
density in the calculation was assumed to be 1/3 of the normal nuclear density and the symmetry energy
co-efficient $\gamma$ was taken to be 25 MeV.
\par 
The calculated primary and secondary fragment yield distributions for the Carbon isotopes
in $^{40}$Ca + $^{58}$Ni, $^{40}$Ar + $^{58}$Ni and $^{40}$Ar + $^{58}$Fe reactions at various excitation
energies are as shown in figure 9. The characteristics of the hot primary fragment yield distribution, shown
in the left column of the figure, change significantly after the secondary de-excitation. The 
primary yield distribution for the three systems, shown by the dotted, dashed and solid curve, for
each excitation energy clearly show isospin effect. The most neutron rich system has the largest
yield for the neutron rich isotopes and the lowest yield for the neutron deficient isotopes. This effect appears 
to decrease with increasing excitation
energy. A similar feature is also observed in the secondary fragment distribution shown in the right column
%Fig 11
    \begin{figure}
    \includegraphics[width=0.47\textwidth,height=0.47\textheight]{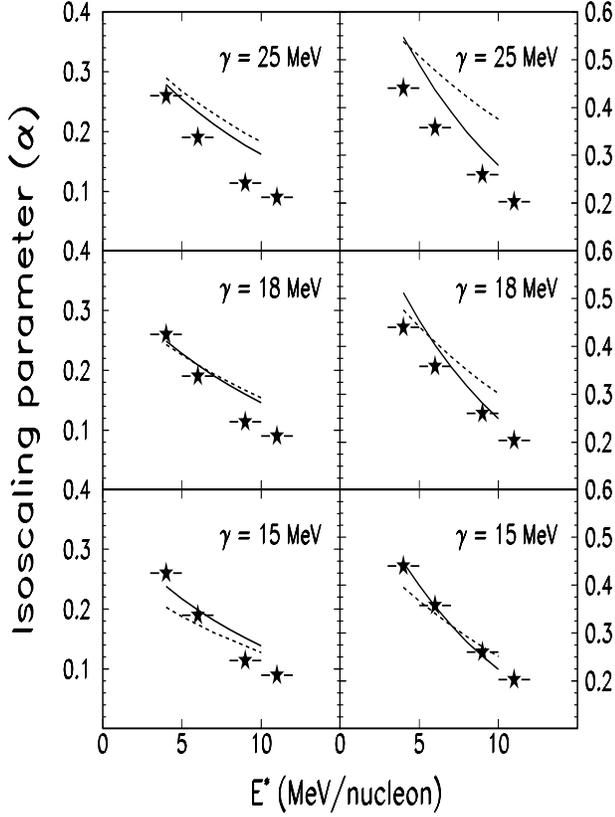}
    \caption{Comparison of the SMM calculated $\alpha$ (lines) with the experimentally determined
      $\alpha$ (symbols) as a function of excitation energy for different values of the symmetry energy co-efficient
      $\gamma$. The dotted lines correspond to the primary fragments and the solid lines to the secondary
      fragments. The left column shows the comparison for the $^{40}$Ar + $^{58}$Ni and $^{40}$Ca +
      $^{58}$Ni pair, and the right column shows the comparison for the $^{40}$Ar + $^{58}$Fe and
      $^{40}$Ca + $^{58}$Ni pair.}
    \end{figure}
%Fig 12
    \begin{figure}
    \includegraphics[width=0.47\textwidth]{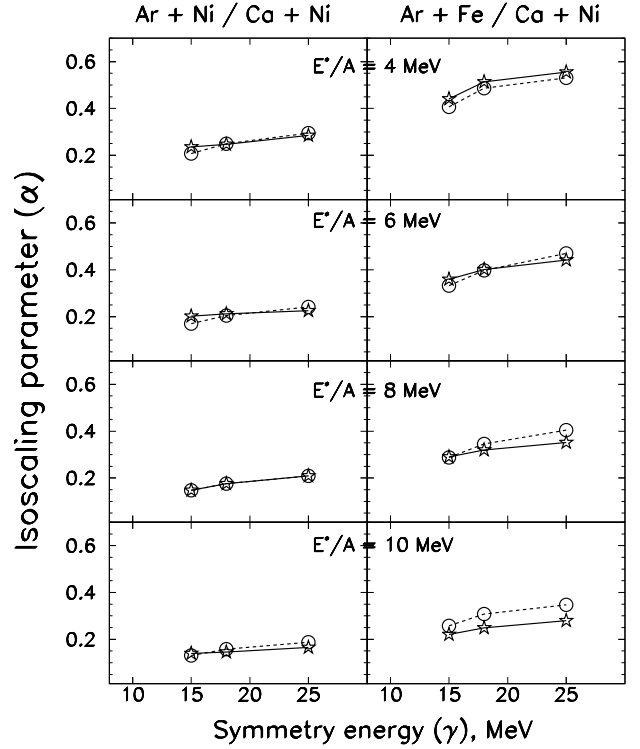}
    \caption{SMM calculated isoscaling parameter $\alpha$ as a function of symmetry energy co-efficient for various 
       excitation energies. The open circles joined by dotted lines correspond to the primary
       fragments and the open stars joined by solid lines to the secondary
      fragments. The left column shows the calculation for $^{40}$Ar + $^{58}$Ni and $^{40}$Ca +
      $^{58}$Ni pair, and the right column for the $^{40}$Ar + $^{58}$Fe and $^{40}$Ca + $^{58}$Ni pair.}
    \end{figure}   
of the figure, though the effect is observed to be weakened significantly. Furthermore, the mean of the 
distribution 
is also observed to decrease along with the width for the secondary fragments. Qualitatively, the SMM 
simulates quite well the overall features of the experimentally 
observed isotopic yield distribution shown in figure 2. The isotopic yield ratios using the primary and the secondary 
fragment distribution from the statistical multifragmentation model are shown in figure 10 for the 
$^{40}$Ar + $^{58}$Ni and $^{40}$Ca + $^{58}$Ni pair of reaction. It is observed that the yield
ratios for both the primary and the secondary distribution obey the isoscaling relation quite well.
Very little difference in the scaling parameter $\alpha$, obtained from the primary and the secondary yield
distribution is observed. Furthermore, the scaling parameter shows a gradual
decrease in its value with increasing excitation energy similar to those observed experimentally and
shown in figure 6.
\par  
Though the overall feature of the scaling parameter calculated from the statistical multifragmentation
model is reproduced quite well, the 
absolute values do not quite agree with the experimentally determined $\alpha$. This is shown in the top 
panel of figure 11, where a comparison is made between the SMM calculated and the experimentally 
observed values of $\alpha$. In the figure, the left column correspond to the  $^{40}$Ar + $^{58}$Ni and 
$^{40}$Ca + $^{58}$Ni pair of reaction and the right to the $^{40}$Ar + $^{58}$Fe and $^{40}$Ca + $^{58}$Ni 
pair. The dotted lines correspond to $\alpha$ calculated from the primary fragment
distribution and the solid lines to those calculated from the secondary fragment distribution. The
symbols correspond to the experimentally determined $\alpha$'s.  It is observed that the experimentally determined 
$\alpha$'s are significantly lower than the calculated values of $\alpha$ using the standard value of the symmetry 
energy co-efficient, $\gamma$ = 25 MeV, for the isolated cold nuclei in their ground states. 
In order to explain the observed dependence of the isoscaling parameter  $\alpha$ on
excitation energy, we varied the $\gamma$ of the hot primary fragment in the SMM input in the range 25 - 15 MeV. As
shown in the center and the bottom panel of the figure, the isoscaling parameter decreases slowly
with decreasing symmetry energy. The experimentally determined $\alpha$ could be reproduced for both 
pairs of systems at all excitation energies using a symmetry energy value of $\gamma$ = 15 MeV.
This value of symmetry energy is significantly lower than the value of $\gamma$ = 25 MeV used for ground
state nuclei.  
\par
Figure 12. shows the calculated $\gamma$ dependence of the isoscaling parameter $\alpha$ from 
the hot primary fragment distribution, and from the cold secondary fragment distribution, for the two pairs of
systems at various excitation energies. The $\alpha$ as a function of symmetry energy for each 
excitation energy and system is observed to decrease with decreasing symmetry energy.  The difference
between the primary fragment $\alpha$ and the secondary fragment $\alpha$ is negligible for the Ar + Ni
and Ca + Ni reaction pair, which has the lowest difference in neutron-to-proton ratio. The difference 
for the Ar + Fe and Ca + Ni pair, which has the highest difference in neutron-to-proton ratio, however is
slightly larger at higher excitation energies. 
%Fig 13
    \begin{figure}
    \includegraphics[width=0.47\textwidth,height=0.47\textheight]{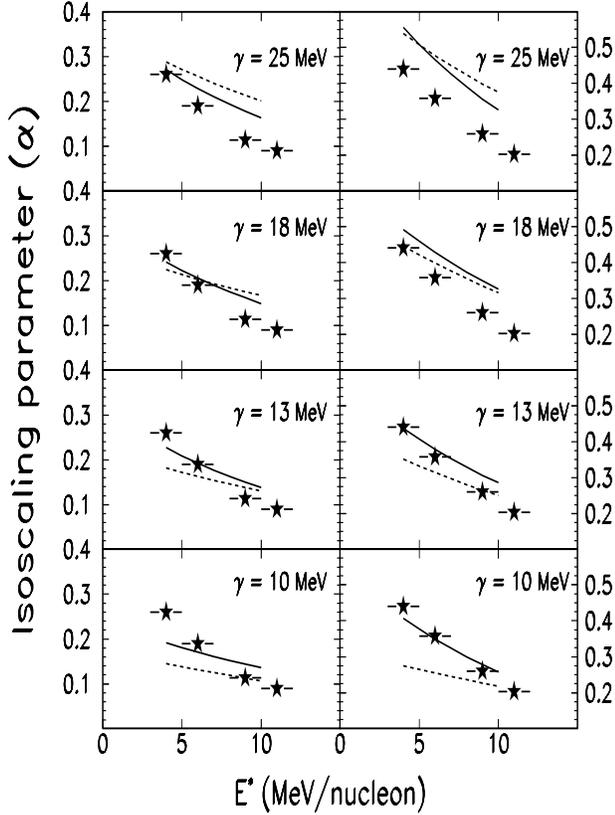}
    \caption{Same as in fig. 11, but with the modified secondary de-excitation with evolving symmetry energy
    co-efficient.}
    \end{figure} 

\subsection{Secondary de-excitation of the fragments with changing symmetry}

In the above described statistical multifragmentation model calculations the masses of the fragments used were 
those of cold isolated nuclei. The fragments in their primary stage are usually hot and the properties of hot
nuclei (i.e., their binding energies and masses) differ from those of cold nuclei. If hot fragments in the freeze-out 
configuration have smaller $\gamma$, their masses at the beginning of the 
secondary de-excitation will be different, and this effect should be taken into account in the evaporation process. 
Recently, Buyukcizmeci {\it {et al.}} \cite{BUY05} adopted a phenomenological approach to estimate the effect of the
symmetry energy evolution during the sequential evaporation. In this approach, they assume liquid drop
masses $m_{A,Z} = m_{ld}(\gamma)$ for the evaporation of the light particles ($n$, $p$, $d$, $t$, $^{3}He$,
$\alpha$), if the internal excitation energy of the fragment is large ($\xi = \beta E^{*}/A > 1$). At lower 
excitation energies ($\xi \leq 1$) they assume a smooth transition to standard experimental masses with shell 
effects($m_{exp}$) using the following dependence,

\begin{equation} 
           m_{A,Z} = m_{ld}(\gamma) \xi + m_{exp} (1 - \xi)
\end{equation}

The excitation energy is determined from the energy balance taking into account the mass $m_{A,Z}$ at the
given excitation. The above corrections were incorporated in the statistical model calculations described 
in the previous section to
study the effect of $\gamma$ during the sequential de-excitation of the hot primary fragments. Figure 13 shows the 
result of the statistical model calculation using the modified secondary de-excitation correction. The isoscaling 
parameter $\alpha$ is
plotted as a function of excitation energy for the two pairs of systems (left and right column).
The top panels show the calculations using symmetry energy value of 25 MeV. As noted in the previous
section, the new calculations are not able to reproduce the experimentally determined alpha for both pairs of systems. 
With decreasing values of the symmetry energy, the calculated $\alpha$ values for the Ar + Ni and Ca + Ni
pair (left column) decrease and
are in better agreement with the experimental values at $\gamma$ = 13 MeV. On the other hand,
the calculated values for the Ar + Fe and Ca + Ni reaction pair (right column) are in good agreement with the experimental
values at $\gamma$ = 10 MeV. In general, one observes that the modified version of secondary de-excitation in SMM leads to a
symmetry energy value of 10 - 13 MeV. This is slightly lower than the value of 15 MeV obtained from the
standard version of the SMM calculation shown in figure 11. 
\par
The dependence of the isoscaling parameter as a function of the symmetry energy for primary and
secondary fragments at various excitation energies are as shown in figure 14. Once again, the difference
between the primary fragment $\alpha$ and the secondary fragment $\alpha$ are extremely small for the Ar + Ni
and Ca + Ni pair of reaction (which has the lowest difference in neutron-to-proton ratio), and only
slightly larger for the Ar + Fe and Ca + Ni pair (which has the highest difference in neutron-to-proton
ratio). However, the main difference between the dependence shown in figure 12 and 14 is the rate at
which the isoscaling parameter $\alpha$ decrease with decreasing symmetry energy.
The decrease is much slower in the calculation  where the symmetry energy dependence of the mass is taken
into account during the secondary de-excitation. The slower decrease in the isoscaling parameter
results in the calculation being able to reproduce the experimental value at a slightly lower value of
symmetry energy. One also observes that the primary fragment $\alpha$ in the modified version of the secondary decay
calculation are consistently smaller than the secondary fragment $\alpha$ at lower symmetry energies. 

%Fig 14
    \begin{figure}
    \includegraphics[width=0.47\textwidth]{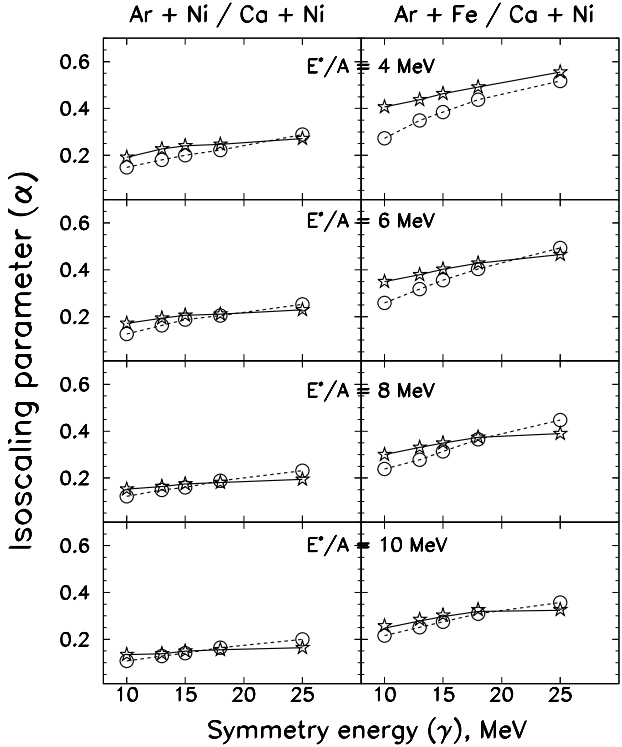}
    \caption{Same as in fig. 12, but with the modified secondary de-excitation with evolving symmetry energy
            co-efficient.}
    \end{figure}
%Fig 15
    \begin{figure}
    \includegraphics[width=0.47\textwidth]{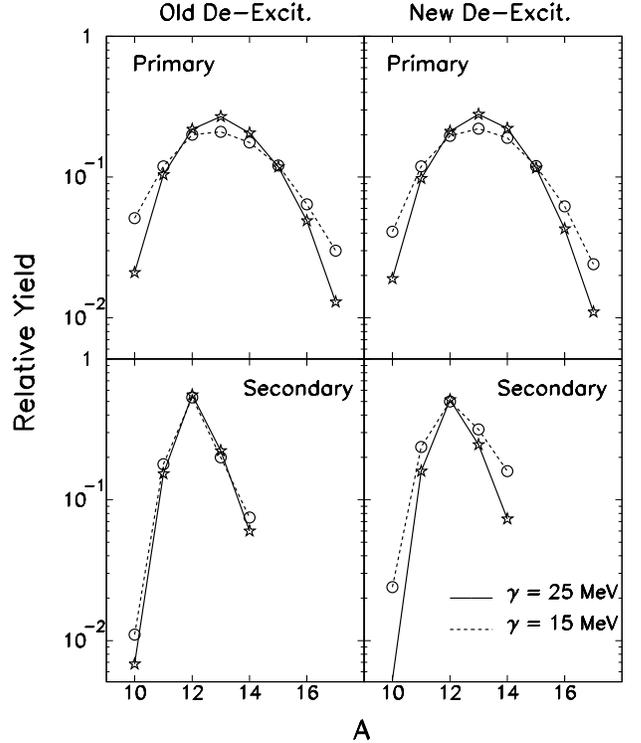}
    \caption{Comparison between the calculated primary (top) and the secondary (bottom) isotopic yield distribution for the
    carbon element in $^{40}$Ar + $^{58}$Fe reaction at $E^{*}$ = 6 MeV/nucleon. The left panels correspond
    to the old and the right to the new de-excitation prescription used in the SMM calculation. The solid and the dashed 
    curves correspond to the calculations using two different values of the symmetry energy.}
    \end{figure}

\par
In order to understand the difference in the symmetry energy, we show in figure 15, the calculated 
isotopic yield distribution for the Carbon element in $^{40}$Ar + $^{58}$Fe reaction at $E^{*}$ = 6 MeV/nucleon. 
The figure shows the primary and the secondary yield distribution using two different
prescriptions for the secondary de-excitations in SMM. The left column in the figure corresponds
to the SMM calculations, where the fragment masses used are those of cold isolated nuclei, and the right
column corresponds to the SMM calculations, where the masses (symmetry energy) evolve with their excitation
energy during secondary de-excitation. 
The panels on the top correspond to the primary yield distribution and those in the bottom to the 
secondary yield distribution. The dotted and the solid curves in each panel corresponds to the
calculations assuming two different values of the symmetry energy, 15 and 25 MeV respectively. From the  
figure, it is evident that there exist a subtle difference between the two final (secondary)
distributions, shown by the dotted and the solid curves in the bottom panels. One also observes that the distribution depends 
on whether the mass (symmetry energy) evolves during the evaporation or not. The SMM calculation with the standard 
de-excitation (i.e the old de-excitation) leads to a narrow final distribution and the isotopes are concentrated close 
to the $\beta$-stability line. The difference in the final yield distributions for $\gamma$ = 15 MeV 
and $\gamma$ = 25 MeV is very small.  This difference is however much more pronounced in the new
de-excitation calculation. The final isotope distributions in this case are considerably
wider, and shifted toward neutron-rich side. The SMM calculation assuming the mass (symmetry energy)
evolution during the evaporation therefore leads to larger yields for neutron rich fragments. 
A similar observation was made by Buyukcizmeci {\it {et al.}} \cite{BUY05} in their calculation of the
primary and secondary fragment isotopic distributions in $^{197}$Au, $^{124}$Sn and $^{124}$La systems. 
By using the experimental masses during the evaporation from the
primary fragments, the emission of the charged particles are suppressed by the binding energy and the Coulomb 
barrier. In the case of small $\gamma$, the binding energy in the beginning of evaporation process essentially 
favors emission of charged particles. When the nucleus has cooled down sufficiently to restore the normal symmetry
energy, the remaining excitation energy is rather low ($E^{*}/A$ $<$ 1 MeV) for the nucleus to evaporate
many neutrons. 
\par
The above comparison of the experimentally observed isoscaling properties with the statistical
multifragmentation model shows that, irrespective of the secondary de-excitation, the final fragment
distribution depends strongly on the available free energy and the strength of the symmetry energy. A 
significantly lower value of the symmetry energy than that assumed for cold isolated nuclei is required 
to explain the isotopic compositions of the fragments produced in multifragmentation reaction. The 
difference between the two kinds of evaporation calculations gives a measure of the uncertainty expected 
in the present analysis. The results above indicate that the properties of nuclei produced at high 
excitation energy, isospin and reduced density could be significantly different from 
those of the cold isolated nuclei. Such information can provide important inputs for the understanding  
of the nuclear composition of supernova matter where hot and neutron rich nuclei are routinely 
produced \cite{BOT04, BOT05}.

\section{Summary and Conclusions}
In summary, we have measured the isotopic yield distribution of the fragments produced in the multifragmentation of 
$^{40}$Ar, $^{40}$Ca + $^{58}$Fe, $^{58}$Ni reactions at 25 - 53 MeV/nucleon. The symmetry energy and the
isoscaling properties of the fragments produced were studied within the framework of statistical multifragmentation 
model. It is observed that the isoscaling 
parameter $\alpha$ for the hot fragments decrease with increasing excitation energy and decreasing 
symmetry energy. The $\alpha$ values increase with increasing difference in the isospin of the fragmenting 
system. Similar behavior is also observed for the cold secondary fragments. 
The sequential decay of the primary fragments to secondary fragments is observed to have very little 
influence on the isoscaling parameter as a function of excitation energies and isospin of the fragmenting system.
The symmetry energy however, strongly influences the isospin properties of the hot fragments. The experimentally
determined scaling parameters could be explained by a symmetry energy that is as low as 10 - 15 MeV,
and significantly lower than that for the normal (cold) nuclei at saturation density. The present results indicate 
that the isospin properties of the fragments produced at high excitation energy and reduced density in 
multifragmentation reaction are sensitive to the symmetry energy.

\section{Acknowledgment}
The authors wish to thank the Cyclotron Institute staff for the excellent beam quality. This work was supported 
in parts by the Robert A. Welch Foundation through grant No. A-1266, and the Department of Energy through grant 
No. DE-FG03-93ER40773.

\end{document}